\documentclass[conference]{IEEEtran}
\IEEEoverridecommandlockouts

\usepackage{array}
\usepackage{stfloats}
\usepackage{url}
\usepackage{verbatim}
\usepackage{algorithm}
\usepackage{subcaption}
\usepackage{hyperref}
\usepackage{cite}
\usepackage{amsmath,amssymb,amsfonts}
\usepackage{algorithmic}
\usepackage{graphicx}
\usepackage{textcomp}
\usepackage{xcolor}
\usepackage{comment}
\usepackage{caption} 
\usepackage{float}
\usepackage{multirow}
\usepackage{makecell}

\usepackage[a4paper, top=0.75in, bottom=1.3in, left=0.6in, right=0.6in]{geometry}

\def\SNR{\mathop{\mathrm{SNR}}} 

\begin{document}

\title{Goal-Oriented Semantic Resource Allocation with Cumulative Prospect Theoretic Agents\\
}

\author{\IEEEauthorblockN{Symeon Vaidanis$^*$,  Photios A. Stavrou$^*$, and Marios Kountouris${^*}^{\dagger}$}\\
$^*$Communication Systems Department, EURECOM, Sophia-Antipolis, France\\
$^{\dagger}$Department of Computer Science and
Artificial Intelligence, University of Granada, Spain\\
Emails: \texttt{\{symeon.vaidanis, fotios.stavrou, marios.kountouris\}@eurecom.fr}}

\maketitle
\thispagestyle{empty}
\pagestyle{empty}

\begin{abstract}
We introduce a resource allocation framework for goal-oriented semantic networks, where participating agents assess system quality through subjective (e.g., context-dependent) perceptions. To accommodate this, our model accounts for agents whose preferences deviate from traditional expected utility theory (EUT), specifically incorporating cumulative prospect theory (CPT) preferences. We develop a comprehensive analytical framework that captures human-centric aspects of decision-making and risky choices under uncertainty, such as risk perception, loss aversion, and perceptual distortions in probability metrics. By identifying essential modifications in traditional resource allocation design principles required for agents with CPT preferences, we showcase the framework's relevance through its application to the problem of power allocation in multi-channel wireless communication systems.
\end{abstract}


\section{Introduction}
Departing from conventional approaches, goal-oriented semantic communication prioritizes the effectiveness of transmitted information by focusing on generating, processing, and delivering content specifically relevant and important to the application or user’s goals \cite{kountouris2021semantics,strinati:2024,stavrou:2023}. This foundational shift in design philosophy minimizes redundant data, significantly enhancing resource and computational efficiency, and will be instrumental in enabling collaborative hyperconnected intelligence and the Internet of Agents.

Goal-oriented semantic communication has unearthed two game-changing principles. First, the semantic of information (SoI), i.e., its value, importance, and utility, is inherently relative and subjective. The significance of information in communication is shaped by the goals, context of use, and user perceptions, which can vary and may be distorted by various factors and interactions with others. Second, the content consumer (observer or decision maker) plays a central role in evaluating the semantic value and perceived utility of information, often through subjective assessment. In simple terms, within a network where not all bits or packets are equally important, the receiver may act both as an objective observer and a subjective perceiver.
In this context, a novel multi-objective network optimization framework that incorporates semantics-aware utilities and subjective perceptions of alternative outcomes, while accommodating diverse risk attitudes under uncertainty, is of cardinal importance. 

While recent studies have investigated various aspects of goal-oriented semantic communications, semantics-aware resource allocation and network optimization remain largely unexplored and challenging. In this paper, we address these aspects leveraging cumulative prospect theory (CPT) \cite{Kahneman_Tversky_1992}, which captures information semantics via risk-sensitive measures, multi-attribute utility functions, and rank-dependent weighting through nonlinear probability transformations. This is a departure from the risk-neutral expected utility theory (EUT) that has dominated conventional network optimization.
Previous studies in (wireless) communication and networking \cite{CPT_Alouini,CPT_ISAC,CPT_UAV,CPT_Protocols,CPT_Gaussian_Channel,CPT_Cloud,CPT_Mobile_Data,CPT_Game_Theory,CPT_Learning} have applied prospect theory \cite{Kahneman_Tversky_1979}, despite its known limitations and inconsistencies. However, to the best of our knowledge, cumulative prospect theory remains largely unexplored in this context. Our work is the first to propose leveraging, adapting, and generalizing CPT for goal-oriented semantic communications.

In this work, we propose a resource allocation framework that incorporates agent interactions, preferences, and decision-making under risk and uncertainty in goal-oriented semantic networks. Our model incorporates agents' subjective, context-dependent perceptions by adopting CPT preferences, diverging from traditional expected utility theory. We develop an analytical framework that integrates human-centric decision-making, such as risk perception, loss aversion, and probability distortions under uncertainty. The interest of this work is two-fold. On one hand, we show how modified and generalized versions of CPT can effectively capture the relativity and subjectivity inherent in goal- and context-specific semantic information and quality perception, as well as risk aversion. On the other hand, we show how semantics-aware metrics may contribute to further generalizations of CPT. Additionally, we provide a generalized, risk-averse utility function to support these advancements. By identifying key adaptations in conventional resource allocation design to accommodate agents with CPT preferences, we show the framework’s effectiveness through its application to power allocation in multi-channel wireless systems.

\section{Preliminaries in Cumulative Prospect Theory}
In this section, we provide a brief overview of the mathematical framework of CPT \cite{Wakker_Book}.

In a nutshell, CPT has the following features that are lacking from EUT: (i) reference
dependence, (ii) loss aversion, (iii) diminishing sensitivity to returns for both, gains
and losses, (iv) probabilistic sensitivity, (v) rank dependence and cumulative probability weighting.

Each agent is associated with a reference point $x_0 \in \mathbb{R}$, a corresponding value function $u: \mathbb{R} \to \mathbb{R}$, and two probability weighting functions $w^{\pm}: [0, 1] \to [0, 1]$, $w^{+}$ for gains and $w^{-}$ for losses, to model uncertainty perception. We say that $(x_0, u, w^{\pm})$ are the CPT features of that agent.

\subsection{Reference Dependence}
Agents perceive value (semantics) and indicate preferences through deviations from an existing \emph{reference point} $x_0$. This reference point may represent an acquired or expected operating level or quantity (e.g., minimum achieved SoI under typical system operation) and may differ across application scenarios. The utility function domain is partitioned into two regions relative to $x_0$: the loss domain $x < x_0$ and the gain domain $x \geq x_0$. In the loss domain, $u(x) < 0, \forall x < x_0$ and $\lim _{x \to x_0^{-}}u(x) \leq 0$, and in the gain domain, $u(x)>0, \forall x > x_0$ and $\lim _{x \to x_0^{+}}u(x) \geq 0$.

\subsection{Utility function and curvature}
Any utility or value function $u(x)$ satisfies the following fundamental properties in the framework of classical CPT in terms of curvature and monotonicity: (i) it is continuous and strictly increasing in $x$; (ii) $u(x_0)=0$\footnote{Generally speaking, $u(x_0)$ can have any value in the interval $\lim _{x \to x_0^{-}}u(x) \leq u(x_0) \leq \lim _{x \to x_0^{+}}u(x)$; in classical CPT, it is assumed to be zero, i.e., $\lim _{x \to x_0^{-}}u(x) = \lim _{x \to x_0^{+}}u(x) = u(x_0) = 0$.}; (iii) $u(x)$ is concave when $x \geq x_0$ and is convex when $x < x_0$ (diminishing marginal utility); (iv) {$u'(x_0^+) < u'(x_0^-)$ (loss aversion).

As $u$ is strictly increasing in the whole domain, $u(x_1) < u(x_2), \forall x_1 < x_2$, hence $0 < \frac{\partial{u}}{\partial{x}}, \forall x < x_0$ and $0 < \frac{\partial{u}}{\partial{x}}, \forall x_0 < x$. 

The curvature of the utility function (marginal utility) characterizes attitudes toward risk (risk aversion) and the agent's sensitivity to varying scales of change within each subdomain. Specifically, in the gain domain, a convex utility function suggests that the agent perceives changes further from $x_0$ more intensely, whereas a concave function indicates that the agent is more sensitive to changes occurring closer to $x_0$. Conversely, in the loss subdomain, the opposite holds. This differs significantly from the typical assumption in EUT that the utility function is concave throughout.
Furthermore, for a symmetric bet—where potential gains and losses are equal - centered around a point $x_1$ within one subdomain and remaining within that subdomain after the outcome, a concave function represents a risk-averse agent, whereas a convex function describes a risk-seeking agent. 

Two widely used CPT utility functions proposed in the literature are the following. The first one is the Kahneman and Tversky utility function \cite{Kahneman_Tversky_1992} given by
\begin{equation}
    u(x) = \left\{
        \begin{array}{ll}
            \left( x - x_0 \right)^\alpha & \textrm{for} ~x \geq x_0 \\
            - \lambda\left( x_0 - x \right)^\beta & \textrm{for} ~x < x_0 \\
        \end{array} 
    \right. 
    \label{eq:KT_utility}
\end{equation}
where $\alpha, \beta \in  (0, 1]$ (for S-shaped utility function) capture the diminishing sensitivity to returns for gains and losses, respectively, and $\lambda >0$ captures the loss aversion. 
Two major limitations are as follows: (i) it is not marginally differentiable at $x_0$, i.e., $\displaystyle \lim_{x \to 0^-} \frac{\partial{u}}{\partial{x}} = -\infty$ and $\displaystyle \lim_{x \to 0^+} \frac{\partial{u}}{\partial{x}} = +\infty$, and (ii) symmetric bet (loss) aversion can only be satisfied if $\alpha = \beta$ and $\lambda > 1$ \cite{Al-Nowaihi}.

Another commonly used utility function, proposed by K\"{o}bberling and Wakker \cite{Kobberling_Wakker}, is given by
\begin{equation}
    u(x) = \left\{
        \begin{array}{ll}
            \lambda_1 \frac{1 - \exp\left( - \alpha (x - x_0)  \right)}{\alpha} & \text{for}~ x_0 \leq x \\
            - \lambda_2 \frac{1 - \exp\left( - \beta (x_0 - x)  \right)}{\beta} & \text{for}~ x < x_0 \\
        \end{array} 
    \right. 
    \label{eq:KW_utility}
\end{equation}
where $\lambda_1$, $\lambda_2$, $\alpha$, and $\beta$ are goal- or agent-specific parameters, all positive to ensure the function’s strictly increasing monotonicity and S-shape. Moreover, the condition for increasing symmetric bet aversion can be satisfied for $\alpha > \beta$ and $\lambda_2 > \lambda_1$ \cite{Dhami_book}. The terms symmetric and increasing symmetric bet aversion will be defined in the following subsection.

\subsection{Loss Aversion}
The CPT utility function exhibits loss aversion, meaning that agents are more sensitive to losses than to equivalent gains. This property is mathematically formulated by requiring that the right-hand marginal derivative of the utility function at the reference point should be smaller than the left-hand derivative, i.e., $u'(x_0^+) < u'(x_0^-)$. Kahneman and Tversky introduced the concept of \emph{symmetric bet aversion}, defined by $u(x_0 + \delta) + u(x_0 -\delta) < 0, \forall \delta>0$ with $u(x_0)=0$ \cite{Kahneman_Tversky_1979}. This criterion implies that all symmetric fair gambles are rejected in favor of maintaining the status quo. They further introduced a stricter version of this criterion, known as \emph{increasing symmetric bet aversion}, expressed as $u(x_0 + \delta_1) + u(x_0 - \delta_1) < u(x_0 + \delta_2) + u(x_0 - \delta_2), \forall 0 < \delta_2 \leq \delta_1$. This can be reformulated as $\frac{\partial{u}}{\partial{x}} | _{x = x_0 + \delta} < \frac{\partial{u}}{\partial{x}} | _{x = x_0 - \delta}, \forall \delta>0$ and $u(x_0) = 0$. This definition implies that the rejection of all symmetric fair gambles is an increasing function (in absolute values) of the step $\delta$.
Neilson extended these definitions by introducing \textit{weak loss aversion}, given by $\frac{u(z)}{z-x_0} < \frac{u(y)}{y-x_0}, \forall y < x_0 < z$ and \emph{strong loss aversion}, given by $\frac{\partial{u}}{\partial{x}} | _{x = z} < \frac{\partial{u}}{\partial{x}} | _{x = y}, \forall y < x_0 < z$ \cite{Neilson}. Strong loss aversion implies weak loss aversion, and the two are equivalent only when the utility function is strictly increasing, twice continuously differentiable, and S-shaped on $\mathbb{R}$.

\subsection{Probability Weighting Function}
A key attribute of CPT is the non-linear probability distortion, in which objective probability is distorted when being perceived by end users according to a probability weighting function (PWF). The PWF typically models uncertainty perception and captures the effect that human agents overweight small probabilities and underweight moderate and high probabilities. The PWFs $w^{\pm}: [0; 1] \to [0; 1]$ are continuous and strictly increasing, with $ w^{\pm}(0) = 0$ and  $w^{\pm}(1) = 1$.

One of the earliest PWFs was proposed in  \cite{Kahneman_Tversky_1992}, and it is defined  as $w(p) \triangleq \frac{p^\delta}{\left( p^\delta + (1-p)^\delta \right)^{\frac{1}{\delta}}}$ with $0 < \delta \leq 1$.
The most widely used probability weighting function is probably the Prelec function \cite{Prelec}, given by 
\begin{equation}
    w(p) = \exp \left( -\gamma \left( -\ln(p) \right)^\theta \right)
    \label{eq:Prelec_PWF}
\end{equation}
where the parameter $0<\theta<1$ controls the curvature of the PWF and the parameter $\gamma>0$ the location of the inflection point relative to the line $w(p) = p$. The effect of these parameters is summarized in Table  \ref{tab:parameters_probability_weighting_function}.
\begin{table}[h]
    \centering
    \begin{tabular}{|c|c|c|c|}
        \hline
         & $0 < \gamma < 1$ & $\gamma = 1$ & $1 < \gamma$\\
        \hline
        $0 < \theta < 1$ & \makecell{inverse S-shape, \\ $\tilde{p} < w(\tilde{p})$}  & \makecell{inverse S-shape, \\ $\tilde{p} = w(\tilde{p})$} & \makecell{inverse S-shape, \\ $\tilde{p} > w(\tilde{p})$} \\
        \hline
        $\theta = 1$ & \makecell{strictly concave, \\ $\tilde{p} < w(\tilde{p})$} & \makecell{linear, \\ $\tilde{p} = w(\tilde{p})$} & \makecell{strictly convex, \\ $\tilde{p} > w(\tilde{p})$} \\
        \hline
        $1 < \theta$ & \makecell{S-shape, \\ $\tilde{p} < w(\tilde{p})$} & \makecell{S-shape, \\ $\tilde{p} = w(\tilde{p})$} & \makecell{S-shape, \\ $\tilde{p} > w(\tilde{p})$} \\
        \hline
    \end{tabular}
    \caption{The effect of the parameters in Prelec function. The inflection point is denoted with $\tilde{p}$.}
    \label{tab:parameters_probability_weighting_function}
\end{table}

\section{Proposed Resource Control Framework}
In this section, we introduce our proposed goal-oriented semantic resource allocation framework. 

Consider a network with a set of agents or users, denoted by $\mathcal{N} = \{1, \ldots, n\}$, and a (finite) set of allocations. Each agent acquires semantic information from $m$ sources (of risk), each with a different importance (semantic payoff) with respect to a reference point. Agent $i$ makes choices or expresses preferences according to CPT, based on a utility function $u_i$ and a PWF $w_i$. This allows agents to subjectively evaluate performance, with distinct, individualized perceptions of risk and semantic value. 

Each agent $i$ is allocated a prospect $\Pi_i=\{(p_i(1),y_i(1)), \ldots, (p_i(k_i),y_i(k_i))\}$ where  $y_i(\ell_i) \geq0, \ell_i \in \mathcal{K}_i$ denotes an outcome (allocation profile), with $\mathcal{K}_i = \{1, \ldots k_i\}$ denoting the set of outcomes for agent $i$, and $p_i(\ell_i), \ell_i \in \mathcal{K}$ is the probability with which outcome $y_i(\ell_i)$ is allocated. The objective is to maximize the aggregate CPT utility for the agents such that a prospect profile $\{\Pi_1, \ldots, \Pi_n\}$ is feasible, i.e., 
\begin{eqnarray*}
\max \sum_{i=1}^nV_i(\Pi_i)\\
\textrm{s.t.} \{\Pi_1, \ldots, \Pi_n\} \in \mathcal{F}
\end{eqnarray*}
where $\mathcal{F}$ is the set of all feasible prospect profiles.
The CPT value of prospect $\Pi_i$ for the $i$-th agent is given by
\begin{equation}
V_i(\Pi_i) = \sum_{\ell_i \in \mathcal{K}_i}d_{\ell_i}(p_i,\pi_i)u(\zeta_i(\ell_i))
\end{equation}
where 
\begin{eqnarray*}
d_{\ell_i}(p_i,\pi_i) &=&w_i(\tilde{p}_i(1)+\ldots+\tilde{p}_i(\ell_i)) \\
&-& w_i(\tilde{p}_i(1)+\ldots+\tilde{p}_i(\ell_i-1))\\
\tilde{p}_i(\ell_i)&=&p_i(\pi_i^{-1}(\ell_i)), \forall \ell_i \in \mathcal{K}_i
\end{eqnarray*}
for permutation $\pi_i:\mathcal{K}_i \to \mathcal{K}_i$ such that allocation $\zeta_i(1) \geq \ldots \geq \zeta_i(\ell_i)$ and $y_i(\ell_i) = \zeta_i(\pi_i(\ell_i))$, $\forall \ell_i \in \mathcal{K}_i$. 

A longer version of this paper will delve into the CPT-based semantic resource allocation framework and the intricate solution to the general optimization problem, which notably is challenging due to its typically nonconvex and non-smooth objective function. 
The formulation presented here offers guidance for optimally allocating risk in mission-critical applications, where both risk and value are perceived subjectively based on the agent's goals and preferences.

For instance, suppose an agent has to allocate its resources (budget) among $m$ sources of risks (e.g., obtaining critical or important data through a certain path or channel), each offering a potential value (payoff) $c_i$. For a set of allocations $\mathcal{A}=\{\alpha_1, \ldots \alpha_m\}$, the agent seeks to maximize its CPT value $V(y_i)$, where $y_i=\sum \alpha_ic_i$. It can be shown that, in the loss subdomain, concentrating risk (i.e., $\alpha_i=1$ for some $i$ and $\alpha_j=0$ for $j\neq i$) is always optimal for a CPT agent, whereas risk diversification ($\alpha_i=1/m, \forall i$) is optimal in the gain subdomain.

\subsection{CPT and the Semantics of Information}
We can show that the SoI metric is highly related to CPT value function. Let $\mathbf{Y}=(Y_1, \ldots, Y_k)^T$ a random vector of $K$ information attributes (random variables) $\{Y_i\}_{i=1}^K$. Given a composite metric $M(\mathbf{y})$ that depends on the random vector $\mathbf{y}$ with multivariate probability density function (PDF) $f_\mathbf{Y}(\mathbf{y})$, we can define the \emph{perceptual utility} of this metric as  
\begin{equation}
\tilde{M}=\int_{\mathbb{R}_+^K}
u(M(\mathbf{y}))\tilde{f}_\mathbf{Y}(\mathbf{y})\mathrm{d}\mathbf{y}= \int_{\mathbb{R}_+^K}
\mathcal{S}(\mathbf{y})\mathrm{d}\mathbf{y},
\end{equation}
where $\tilde{f}_\mathbf{Y}(\mathbf{y})$ is the perceptual multivariate PDF of $\mathbf{Y}$, which is given by
$\tilde{f}_\mathbf{Y}(\mathbf{y})=\frac{\mathrm{d}\tilde{F}_\mathbf{Y}(\mathbf{y})}{\mathrm{d}\mathbf{y}}=\frac{\mathrm{d}w(F_\mathbf{Y}(\mathbf{y}))}{\mathrm{d}\mathbf{y}}$,
and $\mathcal{S}(\mathbf{y})$ represents the SoI metrics \cite{kountouris2021semantics,pappas2021goal}, with the PWF operating as a context-dependent function that adapts qualitative information attributes according to their importance in specific applications. Loosely speaking, identifying the claims that maximize the CPT value is roughly equivalent to determining the optimal allocation that maximizes the SoI.

Here $\tilde{F}_\mathbf{Y}(\mathbf{y})$ denotes the perceptual multivariate CDF, which is typically a nonlinear transformation of the objective multivariate CDF $F_\mathbf{Y}(\mathbf{y})$ by a PWF $w(\cdot)$.

The perceptual utility $\tilde{M}$ can be used to assess the subjective performance associated with the composite metric. Unlike objective performance evaluation metrics, $\tilde{M}$ permits negative values, which reflect a negative subjective perception of the objective performance.

\subsection{Generalized utility functions}
We first propose a generalized form of the K\"{o}bberling and Wakker utility function \cite{Kobberling_Wakker} as a foundation for modeling agents whose behavior may slightly diverge from conventional CPT, enabling the representation of a broader range of risk behaviors.
\begin{equation}
    u(x) = \left\{
        \begin{array}{lll}
            \lambda_1 \frac{\mu_1 - \exp\left( \frac{\alpha}{\gamma_1} \cdot (x - x_0)  \right)}{\alpha} & x_0 \leq x \\
            \lambda_2 \frac{\mu_2 - \exp\left( \frac{\beta}{\gamma_2} \cdot (x - x_0)  \right)}{\beta} & x < x_0 \\
        \end{array} 
    \right. 
    \label{eq:KW_utility_extension}
\end{equation}
where $\alpha$, $\beta$, $\lambda_1$, $\lambda_2$, $\gamma_1$, $\gamma_2$, $\mu_1$ and $\mu_2$ are user specific parameters generally defined on $\mathbb{R}$.

\begin{table}[h]
    \centering
    \begin{tabular}{|c|c|c|}
        \hline
         & Gain & Loss\\
        \hline
        Constant & \makecell{$\gamma_1 \to 0^-$,$0 < \alpha$, \\ $0 < \lambda_1 \cdot \mu_1$} & \makecell{$\gamma_2 \to 0^+$,$0 < \beta$, \\ $\lambda_2 \cdot \mu_2 < 0$} \\
        \hline
        Linear & $\alpha \to 0$,$\frac{\lambda_1}{\gamma_1}<0$ & $\beta \to 0$,$\frac{\lambda_2}{\gamma_2}<0$ \\
        \hline
        Convex & $\frac{\lambda_1}{\gamma_1}<0$,$0<\frac{\alpha}{\gamma_1}$,$\mu_1 \leq 1$ & $\frac{\lambda_2}{\gamma_2}<0$,$0<\frac{\beta}{\gamma_2}$,$1 \leq \mu_2$ \\
        \hline
        Concave & $\frac{\lambda_1}{\gamma_1}<0$,$\frac{\alpha}{\gamma_1}<0$,$1 \leq \mu_1$ & $\frac{\lambda_2}{\gamma_2}<0$,$\frac{\beta}{\gamma_2}<0$,$\mu_2 \leq 1$ \\
        \hline
    \end{tabular}
    \caption{Summary of parameters values for each of the subdomains and the possible shapes of the branch of the utility function}
    \label{tab:parameters_utility_function}
\end{table}

We present the following observations regarding the impact of parameters on the S-shaped utility function\footnote{Since the gain and loss branches of the utility function follow identical mathematical forms and shape conditions, our analysis of the gain branch parameters also holds for the loss branch.}.

\begin{itemize}
    \item Firstly, parameters $\alpha$ and $\gamma$ shape the utility function,  acting as a measure of risk aversion within each subdomain. Specifically, as $\alpha \to 0$, the utility function approaches a linear form with the steepest slope corresponding to a risk-neutral agent. Conversely, as $\alpha$ increases, the maximum value of the function decreases and the shape approximates a step function with a nearly flat slope, representing a fully risk-averse agent. Regarding the parameter $\gamma$, its behavior is opposite to that of $\alpha$: it starts as a step function for small values and transitions into a linear function with diminishing amplitude for large values. This behavior is consistent with the Arrow-Pratt measure of absolute risk aversion \cite{Arrow,Pratt}, which, in the context of our proposed utility function, is given by $\frac{\alpha}{\gamma}$. In summary, by appropriately adjusting the parameters $\alpha$ and $\gamma$, we can achieve the desired shape of the utility function while mitigating the impact of the decreasing amplitude.
    
    \item The parameter $\lambda$ determines the maximum value (saturation level) of the utility function and the slope of the tangent line at the reference point. Additionally, the saturation level can indicate the degree of importance attributed to the agent.
    
    \item The parameter $\mu$ controls the vertical shift of the utility function. Typically, $\mu$ is set to zero in nonlinear cases to ensure continuity at the reference point, where the utility value is zero.
    
\end{itemize}

\section{Case Study: Wireless Power Control with CPT Agents}
In this section, we apply the proposed goal-oriented semantic resource allocation framework to the problem of power allocation with CPT agents. 
We consider the downlink of a wireless system with $N$ orthogonal channels and $N$ agents. Our objective is to determine the optimal power allocation under a total power budget constraint, guided by a semantic quality metric that is subjectively evaluated by each CPT agent. The perceptual metric used here, which also defines the domain, is the signal-to-noise ratio ($\SNR$), given by $\SNR = \frac{P \cdot |h|^2}{N_0}$, where $P$ denotes the allocated transmit power, $h$ is the complex channel coefficient, and $N_0$ is the power of the additive white Gaussian noise (AWGN). Specifically, each CPT agent, assigned to a specific channel\footnote{We assume that the channel assignment is predetermined. The joint assignment and power allocation problem - typically NP-hard - can be approached using a  game-theoretic formulation.}, uses the perceptual subjective utility function defined in (\ref{eq:KW_utility_extension}) with parameters set as $\mu_1 = \mu_2 = 1$ and 
$\alpha$, $\beta$, $\lambda_1$, $\lambda_2 > 0$ and $\gamma_1$, $\gamma_2 < 0$. The reference point is denoted as $\textrm{SNR}_0$. Under the condition $- \frac{\lambda_1}{\gamma_1} < - \frac{\lambda_2}{\gamma_2}$ and with the specified parameter values, (\ref{eq:KW_utility_extension}) satisfies Neilson's definition of strong loss aversion. This utility function is concave over both subdomains, representing an extension of exponential utility to account for loss aversion. The power allocation optimization problem for our setup, is formally expressed as follows:
\begin{equation}
    \begin{aligned}
        \max_{\mathbf{P}} \quad & \sum_{i=1}^{N}{w(p_i) u(SNR(i))}\\
        \textrm{s.t.} \quad & \sum_{i=1}^{N}{P(i)} \leq P_{total}\\
          & 0 \leq P(i) \; \forall i    \\
    \end{aligned}
    \label{eq:Opt_Problem}
\end{equation}
where $w(p_i)$ is the PWF, modeling the $i$-th agent's subjective assessment of probability $p_i$. 
This probability may reflect aspects such as the channel activation the likelihood, success rate, or availability of information flow. We will adopt a divide-and-conquer approach, splitting it into two optimization sub-problems (one for the gain and one for the loss).

The dual Lagrangian problem of \eqref{eq:Opt_Problem} is the following:
\begin{equation}
    \begin{aligned}
        \max_{\mathbf{k},\mu} \min_{\mathbf{P}} \quad & \mathcal{L}(\mathbf{P},\mathbf{k},\mu)\\
    \end{aligned}
\end{equation}
where the $\mathbf{k}$ and $\mu$ are the Lagrangian multipliers and the augmented, unconstrained Lagrangian function is given by 
\begin{equation}
    \begin{aligned}
        \mathcal{L}(\mathbf{P},\mathbf{k},\mu) = - \sum_{i=1}^{N} w(p_i) \cdot {u(SNR(i))} & + \sum_{i=1}^{N}{\mathbf{k}(i) \cdot \left( - P(i) \right)} \\
        + \mu \cdot \left( \sum_{i=1}^{N}{P(i)} - P_{total} \right).
     \end{aligned}
\end{equation}
The original maximization problem (\ref{eq:Opt_Problem}) is concave and satisfies Slater's conditions, resulting in zero duality gap, when $P_{total}$ falls within specific intervals that we provide below. Unavoidably, due to loss aversion, certain values of $P_{total}$ do not fall within these intervals, hence sequential quadratic programming (SQP) can be employed to solve the optimization problem effectively.

The KKT conditions \cite{boyd-book} for the dual problem are applied as follows:
\begin{itemize}
    \item {Stationary condition}: $\frac{\partial\mathcal{L}}{\partial P(i)} = 0~ \forall i \Leftrightarrow$ \\$\mu - \mathbf{k}(i) = w(p_i) \cdot \frac{|h(i)|^2}{N_0} \cdot \frac{\partial u(SNR(i))}{\partial \left( SNR(i) \right)}$.
    \item {Complementary slackness}: $- \mathbf{k}(i) \cdot P(i) = 0, \forall i$, which means that $\mathbf{k}(i) = 0$ if $P(i) > 0$ or $\mathbf{k}(i) > 0$ if $P(i) = 0$.
    \item {Complementary slackness}: $\mu \cdot \left( \sum_{i=1}^{N}{P(i)} - P_{total} \right)=0$, which means that $\mu = 0$ if $\sum_{i=1}^{N}{P(i)} < P_{total}$ or $\mu > 0$ if $\sum_{i=1}^{N}{P(i)} = P_{total}$.
\end{itemize}
Hence, we can combine both conditions as follows:
\begin{equation}
    \mu  = w(p_i) \cdot \frac{|h(i)|^2}{N_0} \cdot \frac{\partial u(\SNR(i))}{\partial \left(\SNR(i) \right)} \; \text{iff} \; P(i) > 0, \forall i.
\end{equation}
The above condition can be solved separately for the gain and loss subdomains, taking into account the utility function’s loss aversion.
\paragraph*{Gain subdomain} 
$P(i) = \frac{N_0}{|h(i)|^2} \cdot \left( \SNR_0 + \frac{\gamma_1}{\alpha} \cdot \ln{\left( - \mu \cdot \frac{1}{w(p_i)} \cdot \frac{\gamma_1}{\lambda_1} \cdot \frac{N_0}{|h(i)|^2} \right)} \right)$ for $\mu \leq - w(p_i) \cdot \frac{\lambda_1}{\gamma_1} \cdot \frac{|h(i)|^2}{N_0}$
\paragraph*{Loss subdomain} $P(i) = \frac{N_0}{|h(i)|^2} \cdot \left( \SNR_0 + \frac{\gamma_2}{\beta} \cdot \ln{\left( - \mu \cdot \frac{1}{w(p_i)} \cdot \frac{\gamma_2}{\lambda_2} \cdot \frac{N_0}{|h(i)|^2} \right)} \right)$ for $\mu > - w(p_i) \cdot \frac{\lambda_2}{\gamma_2} \cdot \frac{|h(i)|^2}{N_0}$.

Therefore, the value of $\mu$ determines whether the $i$-th agent falls within the gain or loss subdomain, as well as its allocated power $P(i)$ and the total allocated power $\sum_{i=1}^{N}{P(i)}$.
\begin{itemize}
    \item If $\mu \leq \hat{\mu}_1 = - \frac{\lambda_1}{\gamma_1} \cdot \frac{\min\{ w(p_i) \cdot |h(i)|^2\}}{N_0}$, then all the agents are in gain subdomain. The total allocated power which is matched to $\hat{\mu}_1$ is defined as $\hat{P}_{total}^{(1)}$.
    \item If $\mu > \hat{\mu}_2 = - \frac{\lambda_2}{\gamma_2} \cdot \frac{\max\{w(p_i) \cdot |h(i)|^2\}}{N_0}$, then all the agents are in loss subdomain. The total allocated power which is matched to $\hat{\mu}_2$ is defined as $\hat{P}_{total}^{(2)}$.
    \item If $\mu \in \mathbb{R}_+^* \backslash A $, $A = A_1 \cup A_2$ where $ A_1 = (0,\hat{\mu}_1] \cup (\hat{\mu}_2,+\infty)$ and $A_2 =  \left\{ \bigcup_{i=1}^{N} \left( - w(p_i) \cdot \frac{\lambda_1}{\gamma_1} \cdot \frac{|h(i)|^2}{N_0}, - w(p_i) \cdot \frac{\lambda_2}{\gamma_2} \cdot \frac{|h(i)|^2}{N_0} \right] \right\}$, some agents will fall into the gain subdomain, while others will belong to the loss subdomain. Given the total power constraint, we can identify the range of values for $\mu$ within which the total power lies as an interior point. To determine the exact value of $\mu$ for a specified total power consumption, a bisection search can be applied within this interval.
\end{itemize}

It should be noted that $\hat{P}_{total}^{(1)}$ is always greater than $\hat{P}_{total}^{(2)}$ due to loss aversion. Moreover, the total power $\sum_{i=1}^{N}{P(i)}$ generally decreases as $\mu$ increases. It is important to emphasize that the region of influence for the $i$-th agent is given by the interval $\left(- w(p_i) \frac{\lambda_1}{\gamma_1} \cdot \frac{|h(i)|^2}{N_0}, - w(p_i)\frac{\lambda_2}{\gamma_2} \cdot \frac{|h(i)|^2}{N_0} \right]$. Specifically, any $j$-th agent within the interval $\left(\frac{w(p_i)}{w(p_j)}\cdot \frac{|h(i)|^2}{N_0}, \frac{w(p_i)}{w(p_j)}\cdot \frac{\lambda_2 \gamma_1}{\gamma_2\lambda_1} \cdot \frac{|h(i)|^2}{N_0} \right]$ falls under the influence of the $i$-th agent. Consequently, the subdomain of the $j$-th agent is determined by the subdomain of the $i$-th agent. Thus, as loss aversion increases, the impact region expands. Additionally, the inverse S-shaped PWF amplifies the influence of agents with lower $p_i$, while reducing the impact of more active agents. Furthermore, if the $j$-th agent is less active than the $i$-th agent (i.e., $\frac{w(p_i)}{w(p_j)} > 1$), the influence region of $i$-th agent expands accordingly.

\section{Simulation Results}
In this section, we present simulation results for the power allocation problem solved above, for $N = 6$ CPT agents. The reference point is set to $\SNR_0 = 7$ dB, spectral density of noise $N_0 = -174\; \text{dBm}/\text{Hz}$, and the parameters of the utility function are $\alpha = 3$, $\beta = 2$, $\lambda_1 = 2$, and $\lambda_2 = 4$. For both subdomains, we set the normalization parameters $\gamma_1$ and $\gamma_2$ to $5$. 
To provide further insight, Fig. \ref{perCDF} shows the perceived cumulative distribution function (CDF) of channel quality under Rayleigh fading, for different parameters of Prelec function.
\begin{figure}
    \centering
    \includegraphics[width=0.87\linewidth]{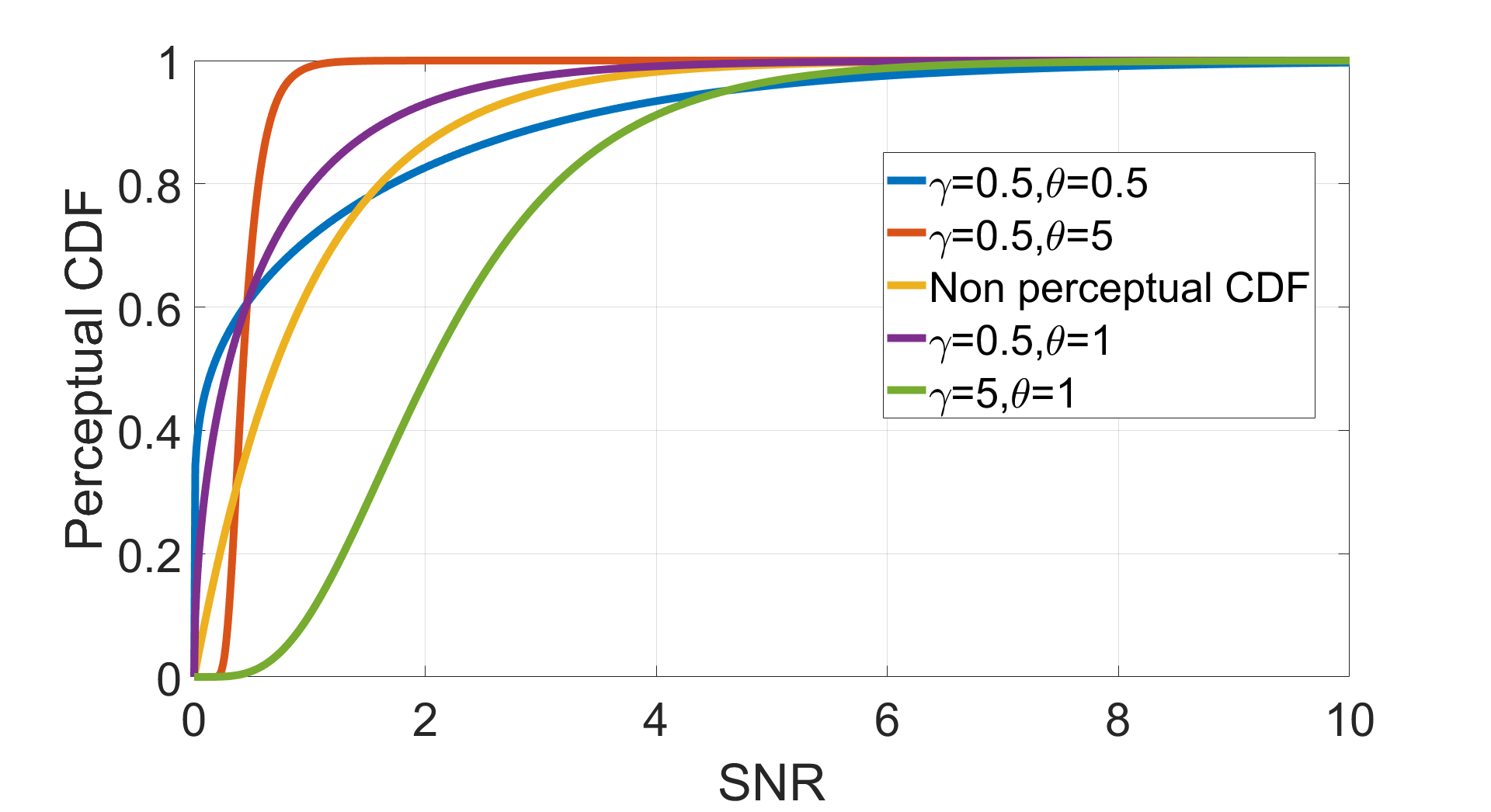}
    \caption{Perceived CDF of channel quality in Rayleigh fading under various parameter configurations.}
    \label{perCDF}
\end{figure}

In Fig. \ref{fig:Allocated_Power}, we plot the allocated power for each agent, ordered by increasing objective unit power channel quality ($|h|^2/N_0$), with $p_i=1, \forall i$. The CPT-based power allocation is compared against both equal power allocation and water-filling strategy. 
With all agents in the gain subdomain in Fig. \ref{fig:Power_Gain}, the power allocation can be understood as an inverse water-filling scheme, primarily due to the utility function's loss aversion. 
As we transition to the intermediate region in Fig. \ref{fig:Power_Intermediate}, where some agents fall within the gain subdomain and others in the loss subdomain, the power allocation starts to transform from an inverse water filling profile of allocation to a more equalized power allocation, going towards an inverse-U shape profile during the passage to the loss subdomain in Fig. \ref{fig:Power_Loss}. We should mention that the inverse-U shape profile is asymmetric. As the total power decreases further and the agents falls deeper to the loss subdomain, the peak of the inverse-U curve shifts to the right.

\begin{figure*}[!t]
    \centering
    \setkeys{Gin}{width=0.26\linewidth} 
    \subfloat[All agents in the gain subdomain.\label{fig:Power_Gain}]{\includegraphics{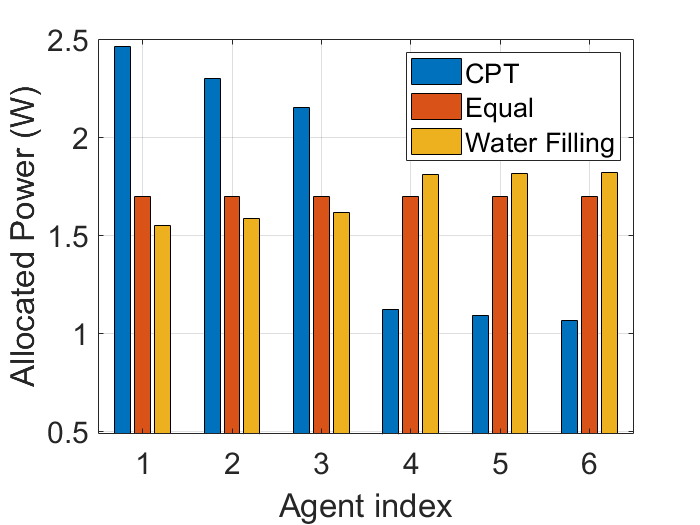} }\hfil
    \subfloat[Intermediate region. \label{fig:Power_Intermediate}]
    {\includegraphics{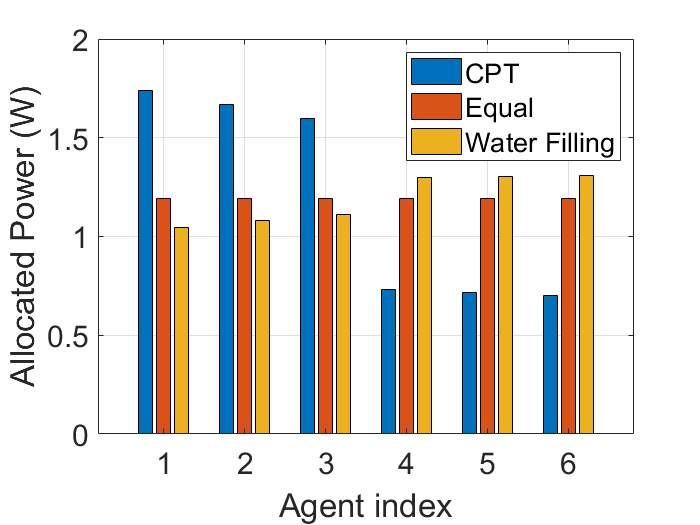} }\hfil
    \subfloat[All agents in the loss subdomain. \label{fig:Power_Loss}]{\includegraphics{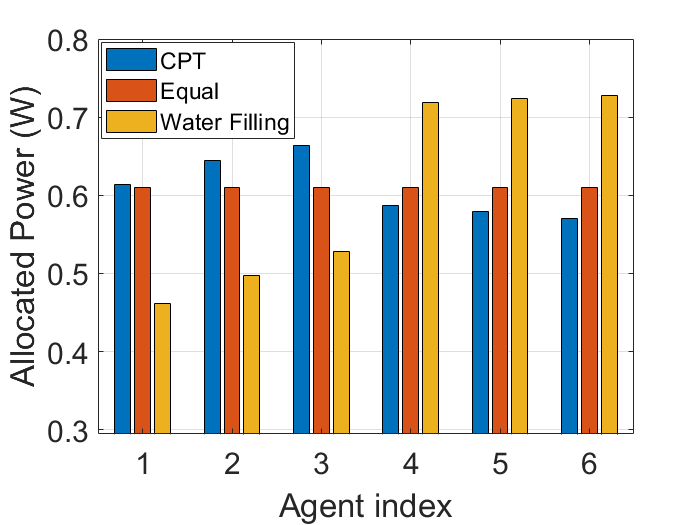} 
    }
    \caption{Optimal power allocation with equal weights $w(p_i) = 1, \forall i$}
    \label{fig:Allocated_Power}
\end{figure*}

In Fig. \ref{fig:Allocated_Power_Weighted}, we plot the power allocation for each agent, arranged by ascending channel quality, with $p_i$ drawn from a uniform distribution over $[0,1]$. When all agents are within the gain subdomain, in Fig. \ref{fig:Power_Weighted_Gain}, the allocation follows an inverse water filling pattern, differing from the unit $w(p)$ case and taking into consideration the weight of each agent. In the intermediate region, shown in Fig. \ref{fig:Power_Weighted_Intermediate}, the inverse water-filling profile begins to decrease, albeit maintaining a weighted by $w(p_i)$ perspective. The most pronounced difference between Fig. \ref{fig:Allocated_Power} and Fig. \ref{fig:Allocated_Power_Weighted} appears in the subfigures \ref{fig:Power_Loss} and \ref{fig:Power_Weighted_Loss}, which depict the scenario where all agents are within the loss subdomain. Comparing the two figures, we observe that, in the first, the allocation profile exhibits an inverse-U shape, whereas in the second, the power allocation profile is strongly influenced by each agent’s probability distortion through $w(p_i)$. 

\begin{figure*}[!t]
    \centering
    \setkeys{Gin}{width=0.26\linewidth} 
    \subfloat[All agents in the gain subdomain.\label{fig:Power_Weighted_Gain}]{\includegraphics{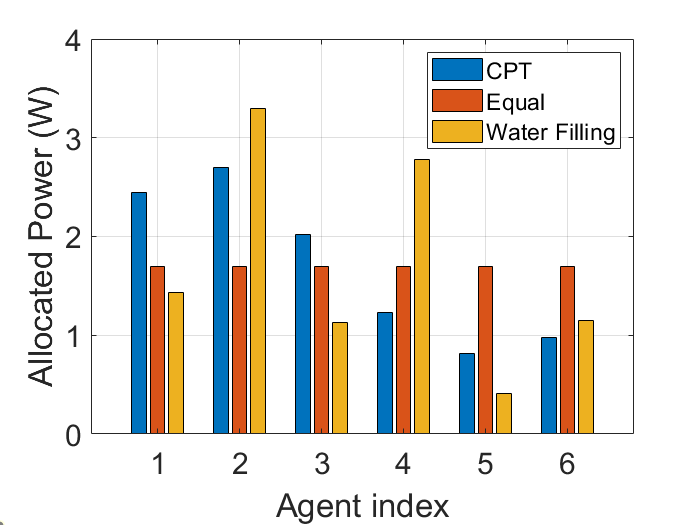} }\hfil
    \subfloat[Intermediate region. \label{fig:Power_Weighted_Intermediate}]
    {\includegraphics{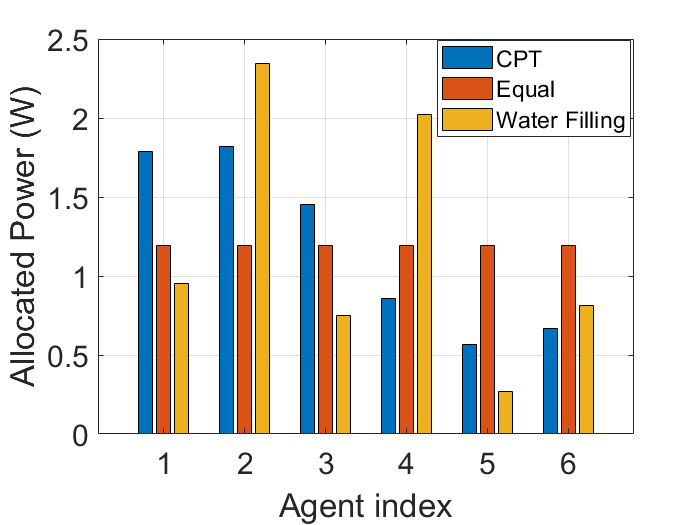} }\hfil
    \subfloat[All agents in the loss subdomain. \label{fig:Power_Weighted_Loss}]{\includegraphics{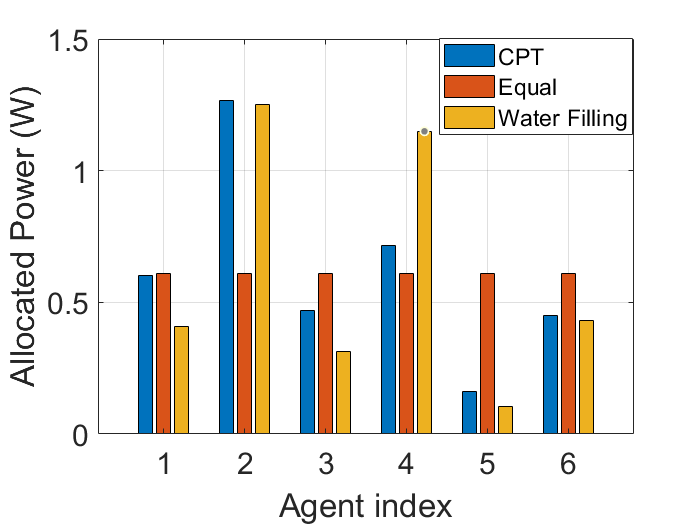} } 
    \caption{Optimal power allocation with unequal weights $w(p_i)$}
    \label{fig:Allocated_Power_Weighted}
\end{figure*}

\section{Conclusions}
In this paper, we presented a novel resource allocation framework for goal-oriented semantic networks, addressing the subjective and context-dependent nature of observer/agent perceptions in evaluating system quality. Leveraging cumulative prospect theory, we account for deviations from traditional expected utility optimization theory, allowing for a more accurate representation of human-centric, risk-averse decision-making under uncertainty. Our analytical framework captured essential aspects such as asymmetric risk perception, loss aversion, and perceptual biases in probability assessment, which are often overlooked in conventional resource allocation approaches. 

\section*{Acknowledgment}
This work is part of a project that has received funding from the European Research Council (ERC) under the EU’s Horizon 2020 research and innovation programme (Grant agreement No. 101003431), from which the work of M. Kountouris is partially supported. The work of S. Vaidanis and P. A. Stavrou is supported by the SNS JU project 6G-GOALS \cite{strinati:2024} under the EU’s Horizon programme Grant Agreement No. 101139232. S. Vaidanis is also supported by the Onassis Foundation - Scholarship ID: F ZU 076-1/2024-2025.

\bibliographystyle{IEEEtran}
\bibliography{my_bibliography}

\end{document}